\documentclass[%
  reprint,
  showpacs,preprintnumbers,
  amsmath,amssymb,
  aps,
  prb,
longbibliography
]{revtex4-1}

\usepackage{graphicx}
\usepackage{dcolumn}
\usepackage{bm}
\usepackage{multirow}
\usepackage{braket}
\usepackage{color}
\usepackage{ulem}

\usepackage{times}
\usepackage{booktabs}

\begin{document}

\preprint{APS/123-QED}

\title{
Meron-antimeron crystals in noncentrosymmetric itinerant magnets on a triangular lattice
}

\author{Satoru Hayami and Ryota Yambe}
\affiliation{
Department of Applied Physics, The University of Tokyo, Tokyo 113-8656, Japan
}
 
\begin{abstract}
Multiple-$Q$ magnetic states often induce nontrivial topological spin textures, such as a skyrmion and a hedgehog. 
We theoretically investigate yet another multiple-$Q$ state with topological defects, a meron-antimeron crystal (MAX), represented by a periodic array of the meron and antimeron with a half-integer skyrmion number. 
Performing simulated annealing for an effective spin model of noncentrosymmetric itinerant magnets on a triangular lattice, we show that rectangular-shaped and triangular-shaped MAXs are stabilized by the interplay between the biquadratic interaction arising from the spin-charge coupling and the Dzyaloshinskii-Moriya interaction arising from the spin-orbit coupling. 
We also discuss the effect of a magnetic field on the triangular MAX, where highly anisotropic responses against a field direction are found. 
In particular, we show that the triangular MAX turns into the skyrmion crystal for the fields along the $y$ and $z$ directions, while it is replaced by another chiral state for the field along the $x$ direction. 
These results would inspire further experimental investigation of the MAXs in itinerant magnets. 
\end{abstract}
\maketitle

\section{Introduction}
\label{sec:Introduction}

Noncentrosymmetric magnets have drawn considerable attention in condensed matter physics, as they exhibit a variety of swirling spin textures~\cite{Bogdanov_PhysRevLett.87.037203,Kenzelmann_PhysRevLett.95.087206,Kimura_PhysRevB.73.220401,nagaosa2013topological,kanazawa2017noncentrosymmetric}. 
The intriguing spin textures have been often induced by the Dzyaloshiskii-Moriya (DM) interaction, which originates from the relativistic spin-orbit coupling in the absence of spatial inversion symmetry~\cite{dzyaloshinsky1958thermodynamic,moriya1960anisotropic}.  
As the DM interaction favors a twisted spin configuration, it can lead to topological swirling spin textures, such as a chiral soliton lattice in a one-dimensional system~\cite{Togawa_PhysRevLett.108.107202,Togawa_PhysRevLett.111.197204,kishine2015theory,togawa2016symmetry,matsumura2017chiral}, a skyrmion crystal (SkX) in a two-dimensional system~\cite{Bogdanov89,Bogdanov94,rossler2006spontaneous,Muhlbauer_2009skyrmion,yu2010real,yu2011near,seki2012observation,Yi_PhysRevB.80.054416,nagaosa2013topological}, and a hedgehog lattice in a three-dimensional system~\cite{Binz_PhysRevB.74.214408,Park_PhysRevB.83.184406,tanigaki2015real,kanazawa2017noncentrosymmetric,Okumura_PhysRevB.101.144416,grytsiuk2020topological,Aoyama_PhysRevB.103.014406}, by combining the effect of magnetic anisotropy and external magnetic field. 
These nontrivial topological spin textures are promising for potential applications to next-generation spintronics devices owing to their large emergent magnetic field~\cite{fert2013skyrmions,romming2013writing,fert2017magnetic,zhang2020skyrmion}.

The above periodic alignment of topological defects is often expressed as a superposition of symmetry-related spin density waves, which is referred as multiple-$Q$ state.  
For example, a hexagonal-shaped SkX is characterized by the triple-$Q$ (3$Q$) state~\cite{rossler2006spontaneous,Muhlbauer_2009skyrmion,yu2010real,yu2011near,seki2012observation,Yi_PhysRevB.80.054416}, while a square-shaped one is characterized by the double-$Q$ (2$Q$) state~\cite{karube2016robust,yu2018transformation,Kurumaji_PhysRevLett.119.237201,khanh2020nanometric,Hayami_PhysRevB.103.024439,Utesov_PhysRevB.103.064414,Wang_PhysRevB.103.104408,Hayami_doi:10.7566/JPSJ.89.103702,Yasui2020imaging,kurumaji2021direct}. 
Besides, multiple-$Q$ states describe various noncollinear and noncoplanar magnetic textures, such as a vortex crystal~\cite{Kamiya_PhysRevX.4.011023,Wang_PhysRevLett.115.107201,Marmorini2014,Hayami_PhysRevB.94.174420,takagi2018multiple,hayami2020phase} and a chiral density wave~\cite{Solenov_PhysRevLett.108.096403,Hayami_PhysRevB.94.024424,Ozawa_doi:10.7566/JPSJ.85.103703,yambe2020double}, and collinear magnetic textures including a bubble crystal~\cite{lin1973bubble,Garel_PhysRevB.26.325,takao1983study,Hayami_PhysRevB.93.184413,seo2021spin}, depending on a way of the superposition of the spin density waves. 
The exploration of new types of multiple-$Q$ states is still active research fields in both theory and experiment. 

The interplay between spin and charge degrees of freedom dubbed the spin-charge coupling in itinerant magnets also leads to the multiple-$Q$ states, as described above\cite{batista2016frustration,hayami2021topological}. 
There, the Ruderman-Kittel-Kasuya-Yosida (RKKY) interaction~\cite{Ruderman,Kasuya,Yosida1957} and higher-order multiple-spin interactions~\cite{Akagi_PhysRevLett.108.096401,Hayami_PhysRevB.90.060402,Ozawa_doi:10.7566/JPSJ.85.103703,Hayami_PhysRevB.95.224424,Nikoli_PhysRevB.103.155151}, which arise from the itinerant nature of electrons, are sources for the instability toward the multiple-$Q$ states even in a centrosymmetric lattice structure~\cite{Ozawa_PhysRevLett.118.147205,Hayami_PhysRevB.99.094420,Hayami_PhysRevB.103.054422}.  
Furthermore, a synergy between the spin-orbit coupling and the spin-charge coupling in noncentrosymmetric itinerant magnets can lead to more exotic multiple-$Q$ states that are not naturally stabilized by each ingredient solely, such as the field-sensitive short-period SkX~\cite{hayami2021field} and the short-period hedgehog lattice~\cite{Okumura_PhysRevB.101.144416}, whose short periodicity might give rise to a giant emergent magnetic field.    

\begin{figure}[t!]
\begin{center}
\includegraphics[width=1.0\hsize]{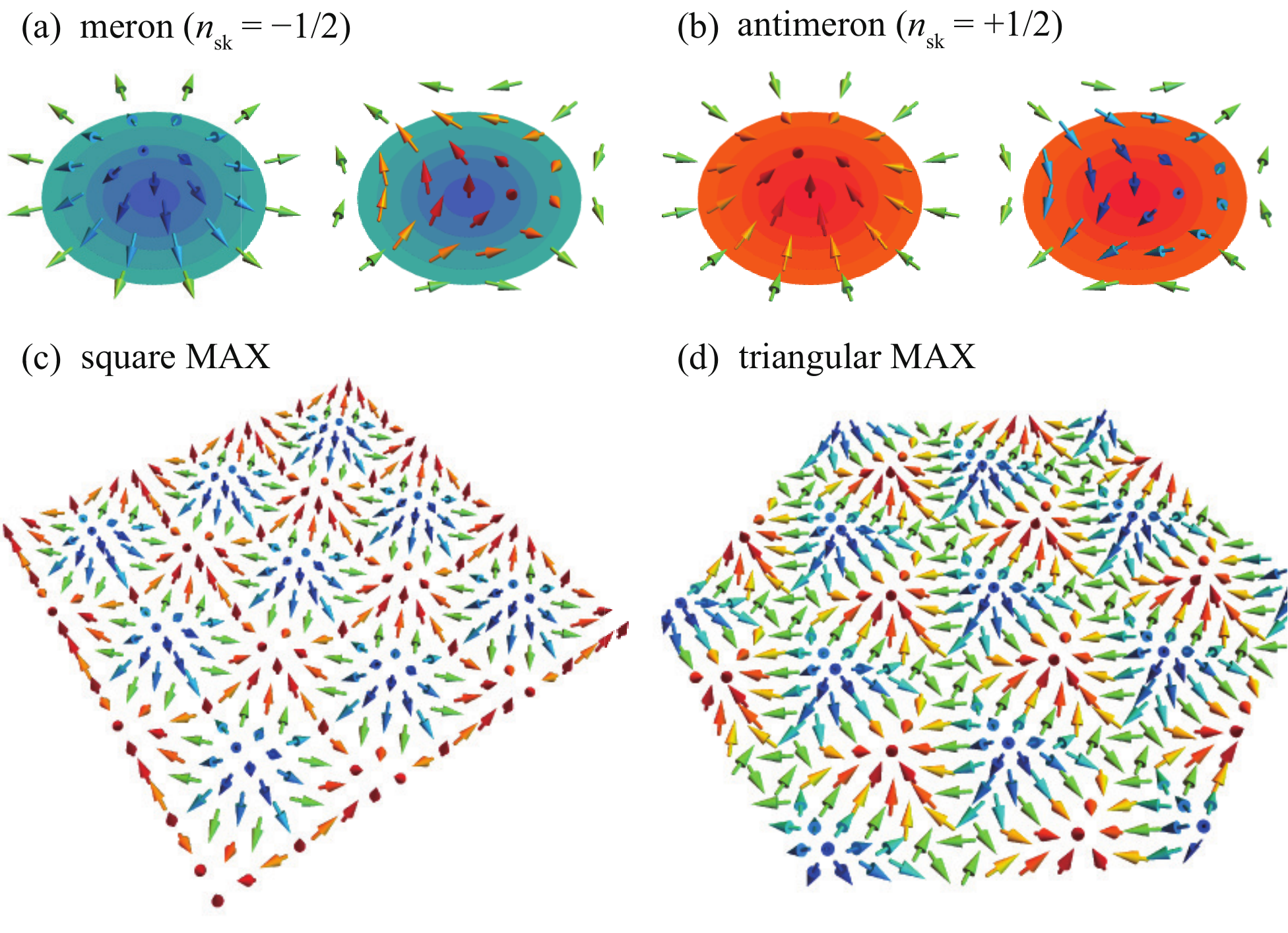} 
\caption{
\label{fig:intro}
(a, b) Schematics of (a) meron and (b) antimeron spin textures with a half skyrmion number $n_{\rm sk}=\mp 1/2$. 
(c, d) MAX on (c) square and (d) triangular lattices. 
}
\end{center}
\end{figure}

Motivated by these studies, we here explore a further possibility of exotic multiple-$Q$ states. 
Especially, we focus on a meron-antimeron crystal (MAX), which is represented by the periodic array of the merons and antimerons. 
Figures~\ref{fig:intro}(a) and \ref{fig:intro}(b) show schematic spin textures of the single meron and antimeron, respectively, where the spins point upwards or downwards in the core region and they are lied in the plane at the perimeter so as to wrap the hemisphere in spin space. 
The meron and antimeron are regarded as a half skyrmion with a half skyrmion number $n_{\rm sk}=-1/2$ and $+1/2$, respectively~\cite{brey1996skyrme,Ezawa_PhysRevB.83.100408,Lin_PhysRevB.91.224407,Tan_PhysRevB.94.014433,Bera_PhysRevResearch.1.033109,Desplat_PhysRevB.99.174409,nych2017spontaneous,gao2019creation}. 
Thus, the topological property of the meron or antimeron is different from that of the skyrmion with a quantized skyrmion number, which results in different physical responses between their periodic structures, the MAX and SkX. 
Although the square MAX in Fig.~\ref{fig:intro}(c) has been observed in chiral magnets Co$_8$Zn$_9$Mn$_3$~\cite{yu2018transformation} and the triangular one in Fig.~\ref{fig:intro}(d) has been suggested in centrosymmetric magnets Gd$_2$PdSi$_3$~\cite{kurumaji2019skyrmion} in experiments, their stabilization mechanism has been theoretically studied mainly for the square MAX~\cite{Hayami_PhysRevLett.121.137202,Hayami_PhysRevB.103.024439,Wang_PhysRevB.103.104408,Utesov_PhysRevB.103.064414}. 
Thus, it is desired to investigate when and how the triangular MAX appears as a stable spin configuration. 

In the present study, we investigate the stability of the triangular MAX in noncentrosymmetric itinerant magnets on a triangular lattice. 
We consider an effective spin model of the noncentrosymmetric itinerant electron model, which includes the symmetric anisotropic RKKY, DM, and biquadratic interactions in momentum space and single-ion magnetic anisotropy. 
By performing simulated annealing for the effective spin model, we show that the interplay between the spin-charge and spin-orbit couplings in the noncentrosymmetric lattice structure is essential to stabilize the triangular MAX.
The triangular MAX is characterized by the 3$Q$ state by superposing three cycloidal waves with different intensities.   
We find that the triangular MAX is transformed to the triangular SkX while increasing the DM interaction, and it is transformed to the rectangular MAX consisting of two cycloidal waves with equal intensities while decreasing the biquadratic interaction. 
We also show that another MAX with two cycloidal modulations with different intensities appears in the region for the large DM and small biquadratic interactions. 
Furthermore, we examine the effect of an external magnetic field on the triangular MAX. 
We obtain the highly anisotropic field-induced transitions in the triangular MAX: 
The SkXs with $n_{\rm sk}=-1$ and $n_{\rm sk}=\pm 1$ are stabilized along the $z$ and $y$ directions, respectively, while another chiral magnetic ordering is induced along the $x$ direction.

The rest of this paper is organized as follows. 
We introduce the effective spin model which includes the effects of the spin-charge and spin-orbit couplings in Sec.~\ref{sec:Model and method}. 
We also outline the numerical method based on simulated annealing. 
In Sec.~\ref{sec:Magnetic phase diagram at zero field}, we present the low-temperature phase diagram at zero magnetic field. 
We show that the interplay between the biquadratic interaction arising from the spin-charge coupling and the DM interaction arising from the spin-orbit coupling is essential to induce the three types of MAXs and the SkX. 
We discuss the effect of the magnetic field on the triangular MAX in Sec.~\ref{sec:Effect of magnetic field}, where different types of chiral states are induced depending on the field direction. 
Section~\ref{sec:Summary} is devoted to a summary.

\section{Model and method}
\label{sec:Model and method}

We start by considering the Kondo lattice model with the relativistic spin-orbit coupling on a noncentrosymmetric triangular lattice with the polar point group $C_\mathrm{6v}$. 
The Kondo lattice model consists of itinerant electrons and localized spins, whose Hamiltonian is given by
\begin{align}
\label{eq:Ham_KLM}
\mathcal{H}^{\rm KLM}= &\sum_{\bm{k}, \sigma} (\varepsilon_{\bm{k}}-\mu) c^{\dagger}_{\bm{k}\sigma}c_{\bm{k}\sigma} 
+   \sum_{\bm{k},\sigma,\sigma'} \bm{g}_{\bm{k}} \cdot c^{\dagger}_{\bm{k}\sigma}\bm{\sigma}_{\sigma \sigma'}c_{\bm{k}\sigma'} \nonumber \\
&+J_{\rm K} \sum_{\bm{k},\bm{q},\sigma,\sigma'}
c^{\dagger}_{\bm{k}\sigma}\bm{\sigma}_{\sigma \sigma'}c_{\bm{k}+\bm{q}\sigma'} \cdot \bm{S}_{\bm{q}},  
\end{align}
where $c^{\dagger}_{\bm{k}\sigma}$ ($c_{\bm{k}\sigma}$) is a creation (annihilation) operator of an itinerant electron at wave vector $\bm{k}$ with spin $\sigma$. 
The first term in Eq.~(\ref{eq:Ham_KLM}) represents the kinetic energy of itinerant electrons, where $\varepsilon_{\bm{k}}$ is the energy dispersion and $\mu$ is the chemical potential. 
The second term in Eq.~(\ref{eq:Ham_KLM}) represents the antisymmetric spin-orbit interaction that originates from the relativistic spin-orbit coupling without inversion symmetry and mirror symmetry with respect to the triangular plane; $\bm{\sigma}=(\sigma^x,\sigma^y,\sigma^z)$ is the vector of Pauli matrices and $\bm{g}_{\bm{k}} =(g_{\bm{k}}^x, g_{\bm{k}}^y, g_{\bm{k}}^z) \propto [-\sqrt{3}\cos (k_x/2)\sin (\sqrt{3}k_y/2), \{2\cos(k_x/2)+\cos(\sqrt{3}k_y/2)\}\sin (k_x/2),0]$ is an antisymmetric vector with respect to the wave vector $\bm{k}$, where we set the lattice constant as unity. 
The third term represents the spin-charge coupling between itinerant electron spins and localized spins via the coupling constant $J_{\rm K}$, where $\bm{S}_{\bm{q}}$ is the Fourier transform of a localized spin $\bm{S}_i$ at site $i$. 
We regard $\bm{S}_i$ as the classical spin with a fixed length $|\bm{S}_i|=1$, where the sign of $J_{\rm K}$ is irrelevant.

By supposing that $J_{\rm K}$ is small compared to the bandwidth of itinerant electrons, the Kondo lattice model in Eq.~(\ref{eq:Ham_KLM}) reduces to an effective spin model, which is given by~\cite{Hayami_PhysRevLett.121.137202,Hayami_PhysRevB.95.224424} 
\begin{align}
\mathcal{H} = \sum_{\nu}
&\left[-J\sum_{\alpha,\beta}I^{\alpha \beta}_{\bm{Q}_{\nu}} S^\alpha_{\bm{Q}_{\nu}} S^\beta_{-\bm{Q}_{\nu}}+\frac{K}{N}\left(\sum_{\alpha,\beta}I^{\alpha \beta}_{\bm{Q}_{\nu}} S^\alpha_{\bm{Q}_{\nu}} S^\beta_{-\bm{Q}_{\nu}}\right)^2 \right.
\nonumber \\
&  \left.-i{\bm D}_\nu\cdot\left({\bm S}_{\bm{Q}_\nu}\times{\bm S}_{-\bm{Q}_{\nu}}\right)
\right]-A\sum_{i}(S^z_i)^2-\sum_{i}\bm{H}\cdot \bm{S}_i. 
\label{eq:Ham}
\end{align}
The term in the square bracket corresponds to the effective interactions obtained by the perturbative expansion in terms of $J_{\rm K}$. 
In the model, we take into account the interaction at the specific wave vectors $\bm{Q}_\nu$ for $\nu=1$-$3$ by supposing that the bare susceptibility of itinerant electrons shows the multiple peaks at $\bm{Q}_\nu$; $\bm{Q}_1=(Q,0)$, $\bm{Q}_2=(-Q/2,\sqrt{3}Q/2)$, and $\bm{Q}_3=(-Q/2,-\sqrt{3}Q/2)$ with $Q=\pi/3$. 
The first term is described by the RKKY interaction and the second term is described by the biquadratic interaction, where $N$ is the number of sites. 
The form factor $I^{\alpha\beta}_{\bm{Q}_{\nu}}$ consists of the isotropic contribution described by the Kronecker delta $\delta^{\alpha\beta}$ and the anisotropic contribution $\tilde{I}^{\alpha\beta}_{\bm{Q}_{\nu}}$, $I^{\alpha\beta}_{\bm{Q}_{\nu}}=\delta^{\alpha\beta}+\tilde{I}^{\alpha\beta}_{\bm{Q}_{\nu}}$.
The anisotropic interaction originates from the bond-dependent magnetic anisotropy to satisfy sixfold rotational symmetry of the triangular lattice, where 
$
-\tilde{I}^{xx}_{\bm{Q}_{1}}=\tilde{I}^{yy}_{\bm{Q}_{1}}=2\tilde{I}^{xx}_{\bm{Q}_{2}}=-2\tilde{I}^{yy}_{\bm{Q}_{2}}=2\tilde{I}^{xy}_{\bm{Q}_{2}}/\sqrt{3}=2\tilde{I}^{yx}_{\bm{Q}_{2}}/\sqrt{3}=2\tilde{I}^{xx}_{\bm{Q}_{3}}=-2\tilde{I}^{yy}_{\bm{Q}_{3}}=-2\tilde{I}^{xy}_{\bm{Q}_{3}}/\sqrt{3}=-2\tilde{I}^{yx}_{\bm{Q}_{3}}/\sqrt{3} \equiv I^{\rm A}$ (the others are ignored for simplicity)~\cite{Hayami_PhysRevB.103.054422,Hirschberger_10.1088/1367-2630/abdef9}. 
$I^{\rm A}<0$ ($I^{\rm A}>0$) combined with the positive isotropic RKKY interaction favors the cycloidal (proper-screw) spiral spin pattern for
each ordering vector. 
A similar bond-dependent magnetic anisotropy has been discussed in terms of the short-ranged one~\cite{Michael_PhysRevB.91.155135,Rousochatzakis2016,amoroso2020spontaneous}. 
The third term represents the DM interaction with the coupling constant $D=|\bm{D}_{\nu}|$. 
Reflecting polar point group symmetry, ${\bm D}_\nu$ is lied along the direction of $\bm{Q}_\nu \times \hat{\bm{z}}$, which favors the cycloidal spiral waves for each ordering vector. 
The coupling constants $J$, $K$, $I^{\rm A}$, and $D$ are proportional to $J^2_{\rm K}$, $J^4_{\rm K}$, $\bm{g}^2_{\bm{k}}$, and $J^2_{\rm K}\bm{g}_{\bm{k}}$, respectively, and also depend on the electronic band structure and chemical potential~\cite{Hayami_PhysRevLett.121.137202}. 
We here take them as phenomenological parameters and ignore the other multiple-spin interactions for simplicity. 
We take $J=1$ as the energy unit of the model in Eq.~(\ref{eq:Ham}). 

The remaining two terms in Eq.~(\ref{eq:Ham}), which are the single-ion anisotropy and the magnetic field coupled to the localized spins, are additionally introduced to the effective spin model. 
For the former, we consider the easy-plane single-ion anisotropy $A<0$, since the easy-axis anisotropy tends to stabilize the SkXs rather than the MAXs~\cite{leonov2015multiply,Lin_PhysRevB.93.064430,Hayami_PhysRevB.93.184413}. 
For the latter, we consider the magnetic field $\bm{H}=(H^x,H^y,H^z)$ for three directions; $\bm{H} \parallel \hat{\bm{x}}$, $\bm{H} \parallel \hat{\bm{y}}$, and $\bm{H} \parallel \hat{\bm{z}}$.   
We fix $A=-0.8$ and $I^{\rm A}=-0.2$ unless otherwise stated.

The instability toward the multiple-$Q$ states in the model in Eq.~(\ref{eq:Ham}) is investigated by carrying out simulated annealing from high temperature. 
We adopt the standard Metropolis local updates in real space in the simulations. 
For a different parameter set of $D$, $K$, and $\bm{H}$, we gradually reduce the temperature of the system with the rate $\alpha=0.99995$-$0.99999$ to the target temperature at $T=0.001$. 
The initial temperature is typically set as $T_0=1$-$5$.
At the final temperature, we perform $10^5$-$10^6$ Monte Carlo sweeps for measurements after equilibration. 
To determine the phase boundary, we also start the simulations from the spin configurations obtained at low temperatures near the phase boundary.
In the following, we present the results for the system with a cluster of $N=96^2$ spins. 

The magnetic phases are identified by computing the spin and scalar chirality configurations. 
The $\alpha=x,y,z$ component of the magnetic moment with wave vector $\bm{q}$ is calculated by 
\begin{align}
\label{eq:mq}
m^\alpha_{\bm{q}} = \sqrt{\frac{S^{\alpha \alpha}_s(\bm{q})}{N}}.
\end{align}
Here, $S^{\alpha \alpha}_s$ is the spin structure factor defined as 
\begin{align}
S^{\alpha \alpha}_s(\bm{q}) = \frac1N \sum_{j,l} \langle S_j^{\alpha} S_l^{\alpha} \rangle e^{i \bm{q} \cdot (\bm{r}_j-\bm{r}_l)},  
\end{align}
where $\bm{r}_j$ is the position vector at site $j$ and $\langle \cdots \rangle$ represents the average over the Monte Carlo samplings. 
We also calculate $(m^{\perp}_{\bm{q}})^2=(m^{x}_{\bm{q}})^2+(m^{y}_{\bm{q}})^2$. 
The $\bm{q}=\bm{0}$ component of $m^\alpha_{\bm{q}}$ stands for the uniform magnetization $m^\alpha_0$. 

The scalar chirality with wave vector $\bm{q}$ is given by 
\begin{align}
\label{eq:chi_q}
&\chi_{\bm{q}}=\sqrt{\frac{S_{\chi}(\bm{q})}{N}}.  
\end{align}
$S_{\chi}(\bm{q})$ denotes the chirality structure factor, which is given by 
\begin{align}
\label{eq:chiralstructurefactor}
S_{\chi}(\bm{q})= \frac{1}{N}\sum_{\mu}\sum_{\bm{R},\bm{R}' \in \mu} \langle \chi_{\bm{R}}
\chi_{\bm{R}'}\rangle e^{i \bm{q}\cdot (\bm{R}-\bm{R}')}, 
\end{align}
where $\mu=(u, d)$ represents upward and downward triangles, respectively, and $\bm{R}$ and $\bm{R}'$ represent the position vectors at the triangle centers. $\chi_{\bm{R}}= \bm{S}_j \cdot (\bm{S}_k \times \bm{S}_l)$ is the local spin chirality at $\bm{R}$, where $j,k$, and $l$ are the sites on the triangle at $\bm{R}$ in the counterclockwise order. 
The uniform component is calculated as $\chi_{0}$, although $\chi_0$ also becomes nonzero for the staggered chirality configuration on upward and downward triangles.

\section{Magnetic phase diagram at zero field}
\label{sec:Magnetic phase diagram at zero field}

\begin{figure}[t!]
\begin{center}
\includegraphics[width=1.0\hsize]{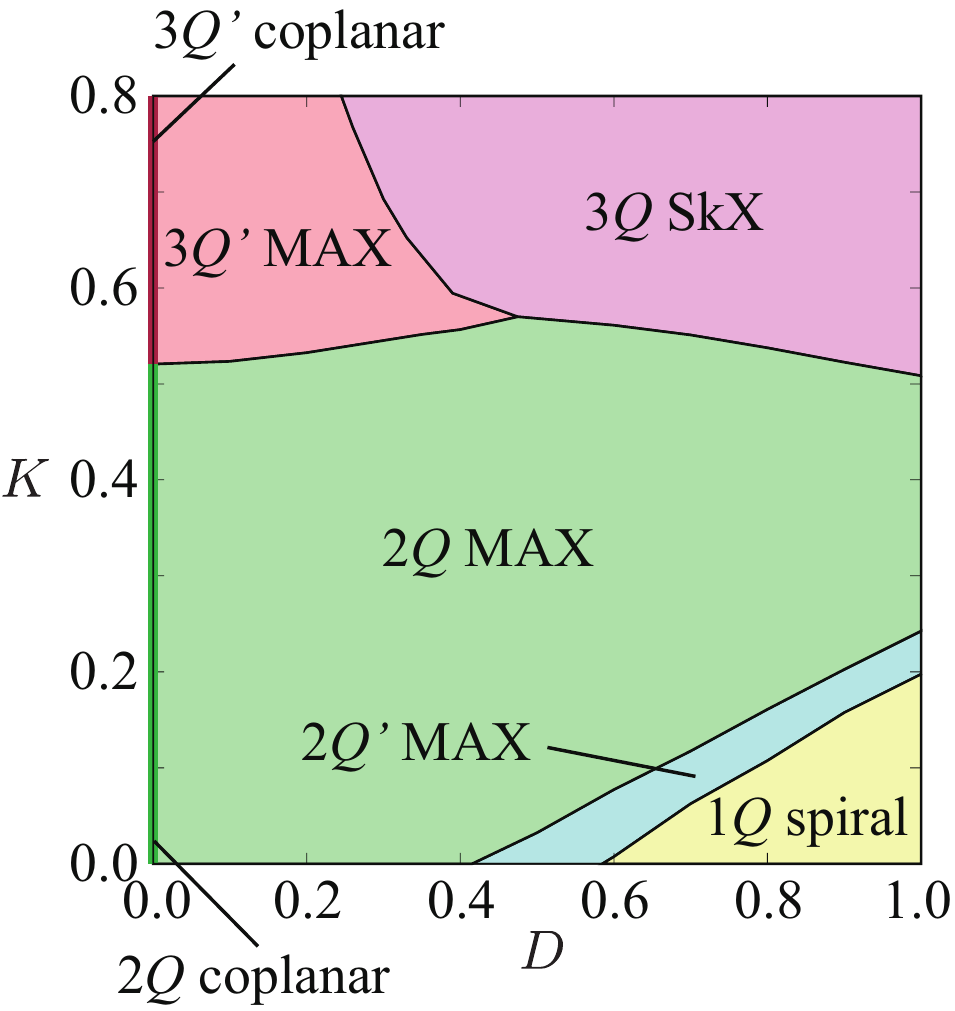} 
\caption{
\label{fig:PD}
Magnetic phase diagram of the model in Eq.~(\ref{eq:Ham}) in the plane of $K$ and $D$ obtained by simulated annealing at $T=0.001$. 
}
\end{center}
\end{figure}

\begin{figure*}[t!]
\begin{center}
\includegraphics[width=0.9\hsize]{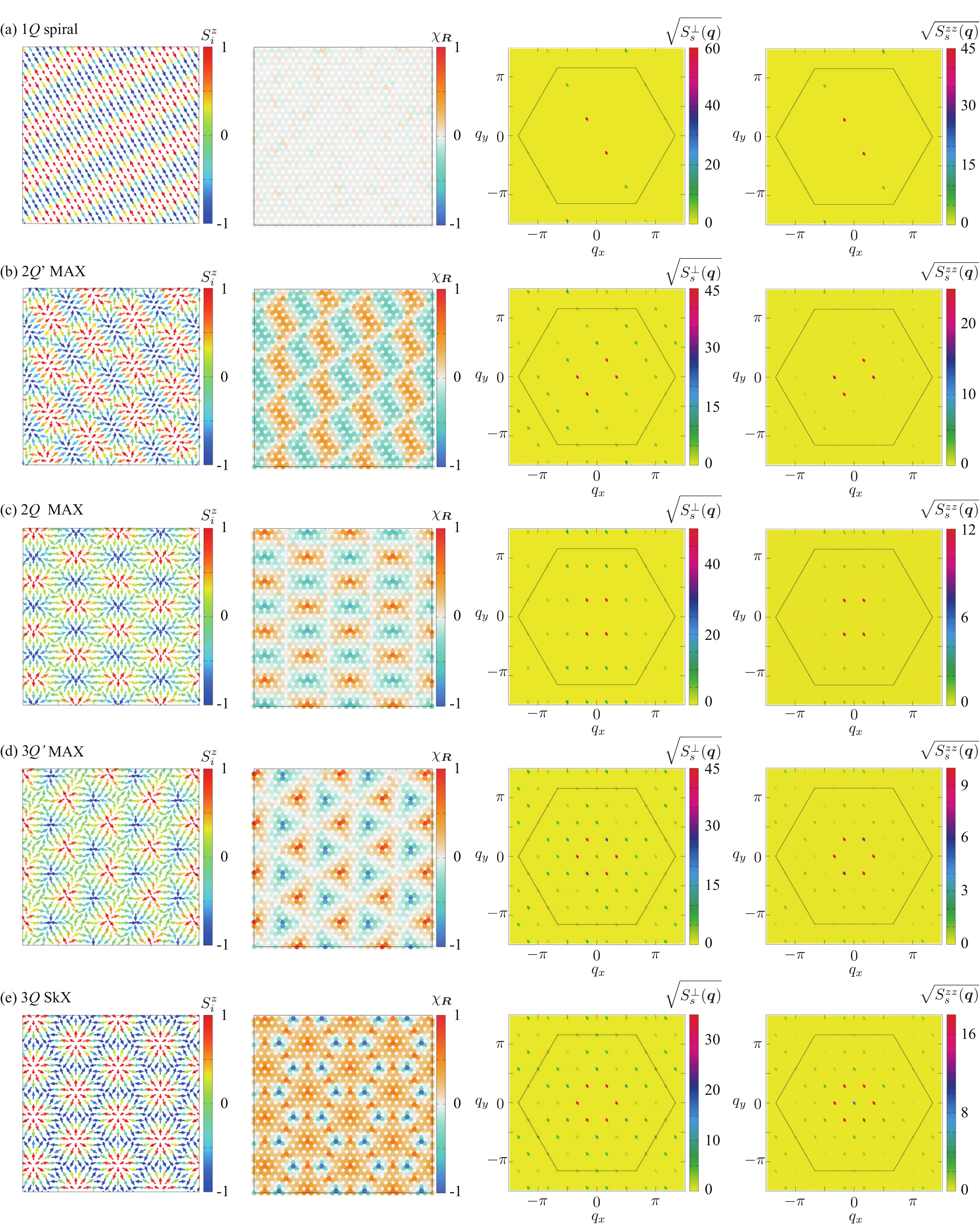} 
\caption{
\label{fig:spin}
 (Left panel) Snapshots of the spin configurations in (a) the 1$Q$ spiral state for $D=1$ and $K=0.1$, (b) 2$Q'$ MAX for $D=0.7$ and $K=0.1$, (c) 2$Q$ MAX for $D=0.2$ and $K=0.2$, (d) 3$Q'$ MAX for $D=0.2$ and $K=0.6$, and (e) 3$Q$ SkX for $D=0.6$ and $K=0.6$. 
The contours show the $z$ component of the spin moment, while the arrows represent the $xy$ components of the spin moment. 
(Middle left panel) Snapshots of the scalar chirality configurations. 
(Middle right and right panels) The square root of the $xy$ and $z$ components of the spin structure factor, respectively. 
In the right two columns, the hexagons with a solid line show the first Brillouin zone.
}
\end{center}
\end{figure*}

In this section, we discuss the instability toward the MAXs in the absence of the magnetic field ($\bm{H}=\bm{0}$). 
Figure~\ref{fig:PD} shows the magnetic phase diagram while varying $D$ and $K$ at a low temperature $T=0.001$ obtained by simulated annealing. 
We find three types of the MAXs denoted as 3$Q'$ MAX, 2$Q$ MAX, and $2Q'$ MAX and the SkX in the presence of $D$ and $K$, as detailed below. 
The spin and chirality configurations in real space and spin structure factors in each state are shown in Fig.~\ref{fig:spin}. 

Let us first discuss the phase diagram at $K=0$, where the model in Eq.~(\ref{eq:Ham}) only includes the bilinear interactions. 
For large $D$, the single-$Q$ (1$Q$) spiral state is stabilized, where the spiral plane is parallel to $\bm{Q}_\nu$, i.e., the cycloidal spiral, so as to gain the energy by the DM interaction.
The spin and chirality configurations of the 1$Q$ spiral state are shown in Fig.~\ref{fig:spin}(a). 
The stripe modulation of the spin is found along the $\bm{Q}_2$ direction, which is characterized by the 1$Q$ peak structure at $\bm{Q}_2$ in the spin structure factor, as shown in the right two panels of Fig.~\ref{fig:spin}(a). 
The appearance of the higher harmonics at $3\bm{Q}_2$ is owing to the elliptical spiral structure with $S^{\perp}_s(\bm{Q}_2)>S^{zz}_s(\bm{Q}_2)$ in the presence of the negative bond-dependent anisotropy and the easy-plane single-ion anisotropy. 
Reflecting the coplanar spin structure, the scalar chirality locally vanishes, as shown in the middle left panel of Fig.~\ref{fig:spin}(a). 

The decrease of $D$ causes the suppression of the $z$ spin component compared to the in-plane spin component, which means that the spiral plane becomes more elliptical.
This imbalance between the $z$ and in-plane spins leads to the enhancement of the higher-harmonic component in the $1Q$ spiral state, which results in the instability towards the multiple-$Q$ states.
As a result, there is a phase transition to the three types of multiple-$Q$ states: 2$Q'$ MAX, 2$Q$ MAX, and 2$Q$ coplanar state, as shown in Fig.~\ref{fig:PD}. 
The small decrease of $D$ induces the 2$Q'$ MAX, which is characterized by the superposition of the 2$Q$ cycloidal waves.
It is noted that the amplitudes of the 2$Q$ peaks are different ($Q'$ means the different amplitudes of the 2$Q$ components), as shown in the right two panels of Fig.~\ref{fig:spin}(b).  
In the real-space picture, this state consists of the two vortices with the same vorticity but the different chirality, as shown in the left two panels of Fig.~\ref{fig:spin}(b). 
As the (anti)vortices with the negative (positive) chirality have (anti)meron-like spin textures, this spin configuration is regarded as the periodic alignment of the meron-antimeron pairs. 
This is why we call this state the 2$Q'$ MAX. 
The meron-antimeron pairs are aligned in a rectangle way owing to the double-$Q$ nature on the triangular lattice.
The positive and negative contributions to the uniform scalar chirality are canceled out in the 2$Q'$ MAX.

While further decreasing $D$, the amplitudes of the 2$Q$ peaks become equivalent, where we call the state the 2$Q$ MAX. 
The spin and chirality textures in the 2$Q$ MAX resemble those in the 2$Q'$ MAX, as shown in Figs.~\ref{fig:spin}(b) and \ref{fig:spin}(c). 
When $D=0$, the $z$ component of the spin vanishes, and then, the 2$Q$ coplanar state consisting of two in-plane sinusoidal waves appears in the phase diagram owing to the bond-dependent anisotropy. 

Next, we turn to the phase diagram for $K>0$. 
By introducing $K$, the multiple-$Q$ nature is enhanced, as shown in Fig.~\ref{fig:PD}; the 1$Q$ spiral state changes into the 2$Q'$ MAX and the 2$Q'$ MAX changes into the 2$Q$ MAX.  
For $0.25 \lesssim K \lesssim 0.5$, only the 2$Q$ MAX appears in the phase diagram. 
The instabilities toward the 3$Q$ states occur for $K\gtrsim 0.5$. 
For small $D$, the 2$Q$ MAX is replaced by the 3$Q'$ MAX. 
The 3$Q'$ MAX is characterized by the 3$Q$ peaks in both $xy$ and $z$ spin components, although their amplitudes are different with each other, as shown in the right two panels of Fig.~\ref{fig:spin}(d). 
Accordingly, the vortices with positive scalar chirality (antimeron) and the those with negative scalar chirality (meron) form a distorted triangular lattice shown in the left two panels of Fig.~\ref{fig:spin}(d), where the uniform scalar chirality is zero. 
It is noted that the different amplitudes at $\bm{Q}_1$-$\bm{Q}_3$ in the MAX induce the uniform in-plane magnetization along the $\bm{Q}_3$ direction for $|\bm{m}_{\bm{Q}_3}| < |\bm{m}_{\bm{Q}_1}|, |\bm{m}_{\bm{Q}_2}|$. 
For $D=0$, the 3$Q'$ MAX is replaced by the 3$Q'$ coplanar state without the $z$-spin component~\cite{takagi2018multiple}.

Meanwhile, the 3$Q$ SkX appears upon increasing $K$ for large $D$. 
This state has a 3$Q$ peak structure with equal intensity, and the skyrmion core forms the triangular lattice, as shown in Fig.~\ref{fig:spin}(e). 
The real-space spin configuration is represented by the periodic alignment of two types of vortices with $S_i^z \simeq \pm 1$, where the number of the voritces with $S_i^z \simeq -1$ is twice as those with $S_i^z \simeq +1$. 
The snapshot of the 3$Q$ SkX obtained by simulated annealing in the left two panels of Fig.~\ref{fig:spin}(e) shows a negative out-of-plane magnetization and a positive uniform scalar chirality.
Although the spin texture in the 3$Q$ SkX is similar to that in the external magnetic field, the obtained state at zero field is energetically degenerate with the one with a positive magnetization and a negative scalar chirality. 
The degeneracy is lifted by the external magnetic field, as will be discussed in Sec.~\ref{sec:z}. 

Both the 3$Q$ SkX and 3$Q'$ MAX are characterized by the superposition of the three cycloidal waves at $\bm{Q}_1$, $\bm{Q}_2$, and $\bm{Q}_3$. 
The main difference between them can be described by the relative phase degree of freedom among the constituent waves~\cite{kurumaji2019skyrmion,hayami2020phase}. 
In the 3$Q$ SkX with nonzero $\chi_0$, the sum of the phase in the three spiral waves becomes zero, $\sum_\nu \theta_{\nu} =0$, where the spiral spin texture is given by $\bm{S}_i = [\cos (\bm{Q}_1\cdot \bm{r}_i+\theta_1), 0,  \sin (\bm{Q}_1\cdot \bm{r}_i+\theta_1)]$ along the $\bm{Q}_{1}$ direction for instance, while that leads to $\sum_\nu \theta_{\nu} =\pi/2$ in the 3$Q'$ MAX without $\chi_0$~\cite{hayami2020phase}. 
Thus, the present result for the phase diagram in Fig.~\ref{fig:PD} implies that the change of $D$ in the presence of large $K$ and $A$ can drive the phase shift in the triple-$Q$ states, which results in the topological phase transition between the MAX and the SkX. 

Another difference between the MAX and the SkX appears in the direction of the spontaneous magnetization at zero field; the SkX has the out-of-plane magnetization, while the MAX shows the in-plane magnetization [see the right two panels of Figs.~\ref{fig:spin}(d) and \ref{fig:spin}(e)].
The result indicates that the easy-axis (easy-plane) anisotropy and/or out-of-plane (in-plane) magnetic field tends to favor the SkX (MAX), which is consistent with the previous studies~\cite{leonov2015multiply,Lin_PhysRevB.93.064430,Hayami_PhysRevB.93.184413}. 
Meanwhile, it is noted that the $3Q$ MAX with the same amplitude at $\bm{Q}_1$-$\bm{Q}_3$ does not exhibit a uniform magnetization in any directions, which implies that the energy gain by the uniaxial anisotropy and/or field is smaller than the 3$Q'$ MAX. 
This is why the 3$Q'$ MAX is realized in the phase diagram instead of the 3$Q$ MAX.

\begin{figure}[t!]
\begin{center}
\includegraphics[width=0.95\hsize]{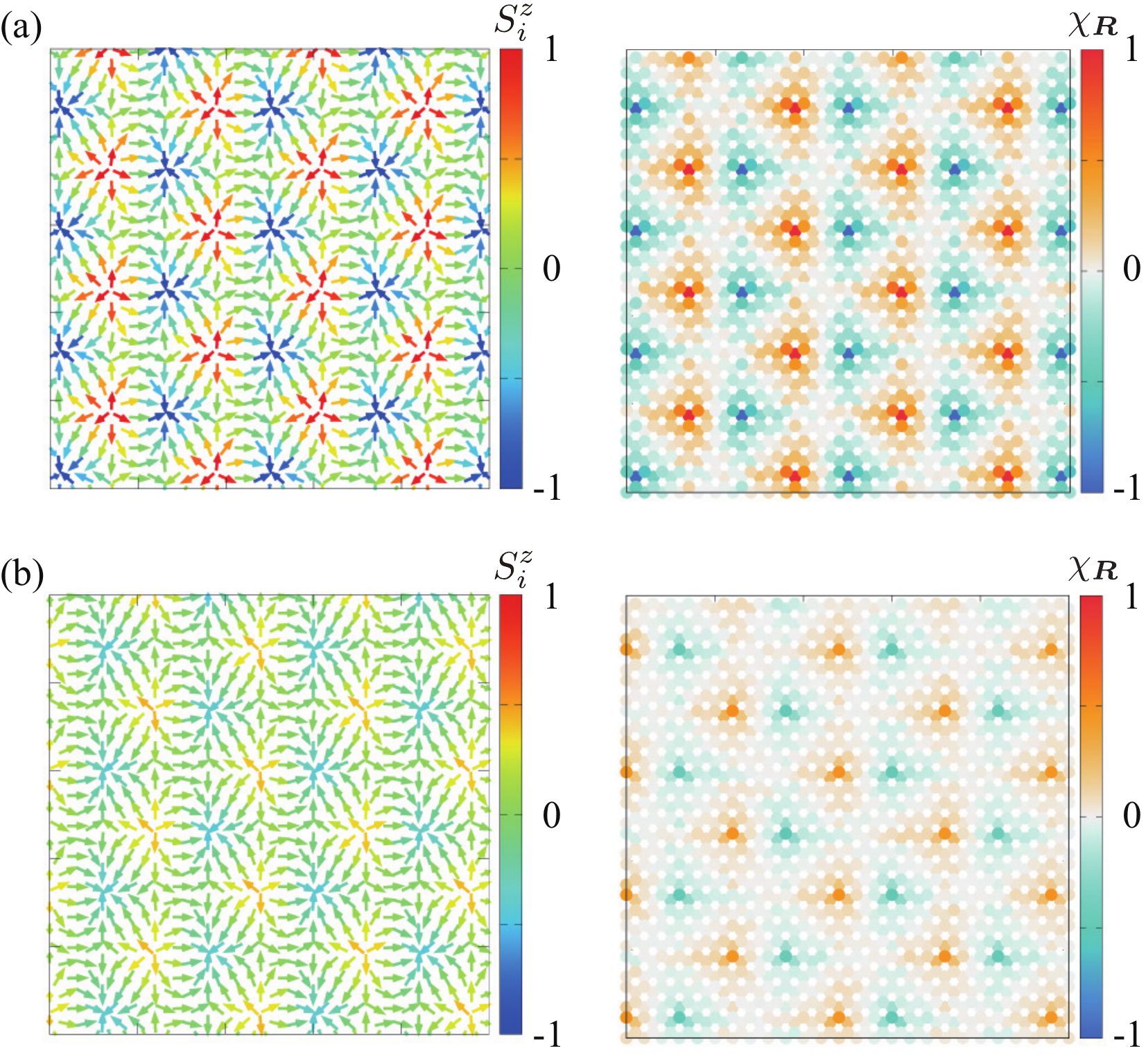} 
\caption{
\label{fig:spindiff}
 (Left panel) Snapshots of the spin configurations for (a) $A=-0.8$ and (b) $A=-2$ at $D=0.2$, $K=0.6$, and $I^{\rm A}=0$. 
The contours show the $z$ component of the spin moment, while the arrows represent the $xy$ components of the spin moment. 
(Right panel) Snapshots of the scalar chirality configurations. 
}
\end{center}
\end{figure}

The 3$Q'$ MAX is robustly stabilized by changing $A$ and $I^{\rm A}$. 
When we fix $I^{\rm A}=-0.2$, the 3$Q'$ MAX appears for $-2 \le A \lesssim -0.5$, where we change $A$ from $-2$ to $0$. 
Meanwhile, we obtain the 3$Q'$ MAX as the ground state for $-0.4 \lesssim I^{\rm A} \le 0$ when we fix $A=-0.8$, where we change $I^{\rm A}$ from $-0.5$ to $0$. 
We show the spin and chirality configurations of the 3$Q'$ MAX for $A=-0.8$ in Fig.~\ref{fig:spindiff}(a) and $A=-2$ in Fig.~\ref{fig:spindiff}(b) at $D=0.2$, $K=0.6$, and $I^{\rm A}=0$. 
This result indicates that the easy-plane anisotropy is also a key factor in stabilizing the 3$Q'$ MAX, whereas the bond-dependent anisotropy is not necessary. 

\section{Effect of magnetic field}
\label{sec:Effect of magnetic field}

In this section, we examine the effect of the external magnetic field on the 3$Q'$ MAX, which is stabilized at zero field. 
We discuss the result along the $z$-field direction in Sec.~\ref{sec:z}, that along the $x$-field direction in Sec.~\ref{sec:x}, and that along the $y$-field direction in Sec.~\ref{sec:y}. 
Hereafter, we fix $D=0.2$ and $K=0.6$. 

\subsection{Field along the $z$ direction}
\label{sec:z}

\begin{figure}[t!]
\begin{center}
\includegraphics[width=0.9\hsize]{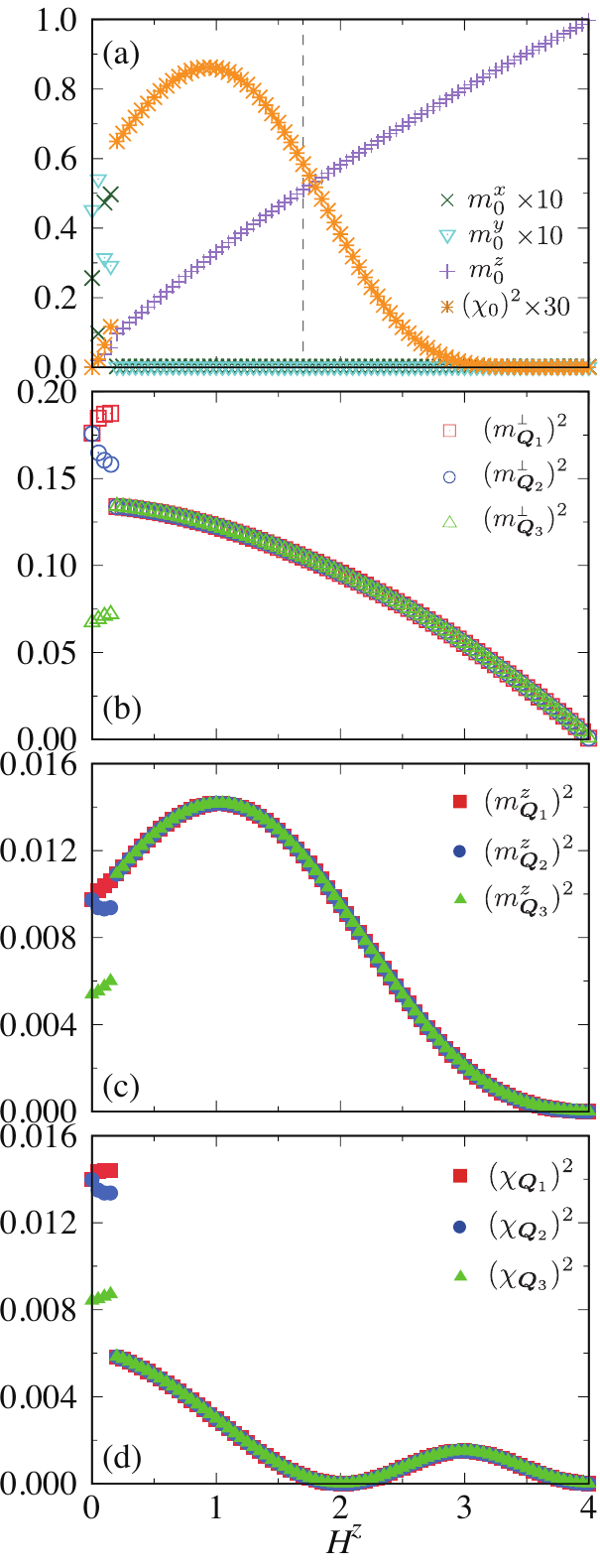} 
\caption{
\label{fig:Hz}
$H^z$ dependence of (a) $\bm{m}_0=(m^x_0,m^y_0,m^z_0)$ and $(\chi_0)^2$, (b) $(m^{\perp}_{\bm{Q}_\nu})^2$, (c) $(m^{z}_{\bm{Q}_\nu})^2$, and (d) $ (\chi_{\bm{Q}_\nu})^2$ for $D=0.2$ and $K=0.6$. 
The dashed line in (a) represents the boundary between the skyrmion number of $-1$ and 0.
}
\end{center}
\end{figure}

\begin{figure}[t!]
\begin{center}
\includegraphics[width=0.95\hsize]{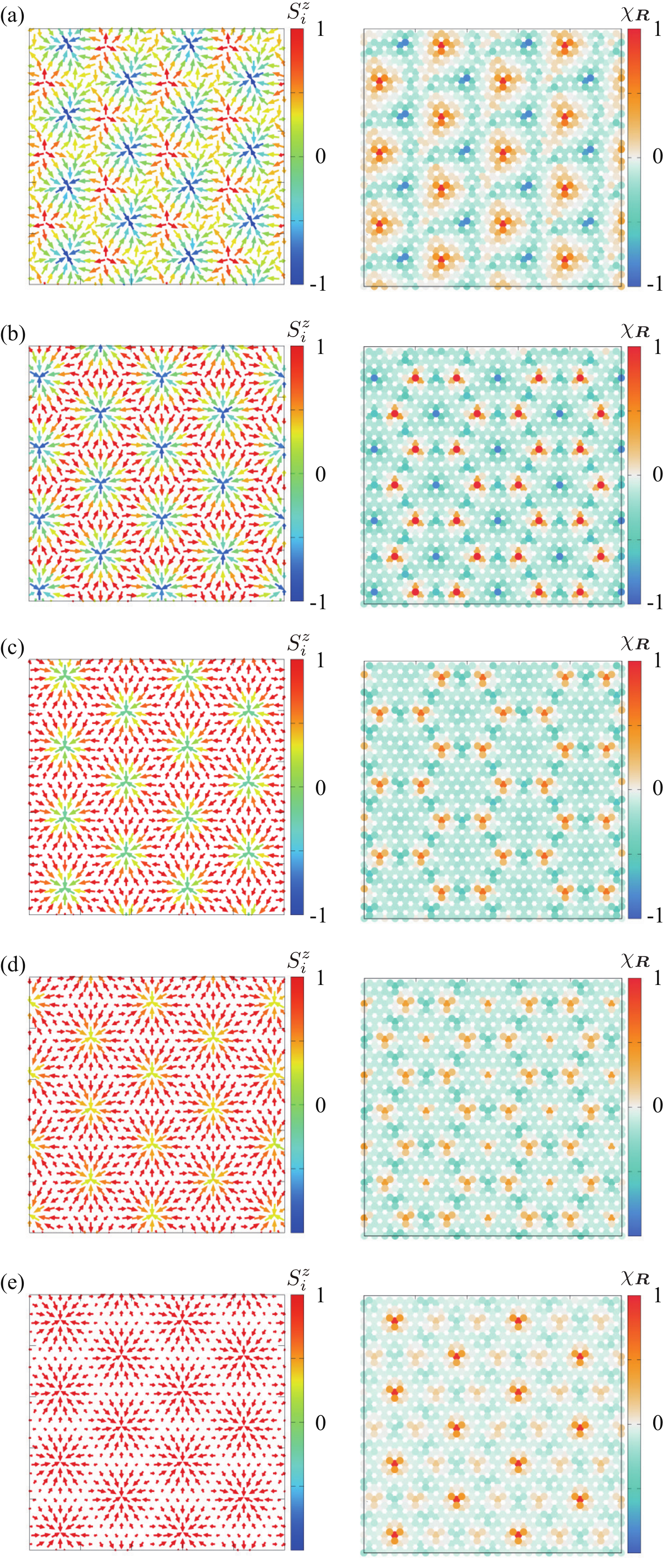} 
\caption{
\label{fig:spin_Hz}
 (Left panel) Snapshots of the spin configurations for (a) $H^z=0.15$, (b) $H^z=0.7$, (c) $H^z=1.5$, (d) $H^z=2$, and (e) $H^z=3$ at $D=0.2$ and $K=0.6$. 
The contours show the $z$ component of the spin moment, while the arrows represent the $xy$ components of the spin moment. 
(Right panel) Snapshots of the scalar chirality configurations. 
}
\end{center}
\end{figure}

Figure~\ref{fig:Hz} shows the result in the presence of the magnetic field along the $z$ direction, $\bm{H}=(0,0,H^z)$. 
By introducing $H^z$, the 3$Q'$ MAX exhibits a uniform scalar chirality $\chi_0$ as well as the uniform magnetization $m_0^z$, as shown in Fig.~\ref{fig:Hz}(a). 
This is because the regions of $S_i^{z}>0$ extend, whereas those of $S_i^{z}<0$ shrink while increasing $H^z$, as found in the snapshot in Fig.~\ref{fig:spin_Hz}(a). 
This state has a quantized skyrmion number of $-1$. 
Thus, the 3$Q'$ MAX under the $z$-field is regarded as the 3$Q'$ SkX with $n_{\rm sk}=-1$

While increasing $H^z$, there is a phase transition to the 3$Q$ SkX at $H^z \simeq 0.2$ with a discontinuous change of the spin- and chirality-related quantities.
Although the value of $\chi_0$ jumps in the transition, the skyrmion number remains $-1$.  
The 3$Q$ SkX is characterized by equal weights of spin and chirality at $\bm{Q}_1$-$\bm{Q}_3$, as shown in Figs.~\ref{fig:Hz}(b)-\ref{fig:Hz}(d). 
This transition might be attributed to the different favorable direction of the magnetization in the MAX and the SkX, as discussed above.
The spin and chirality configurations obtained by simulated annealing are shown in Fig.~\ref{fig:spin_Hz}(b) at $H^z=0.7$, which is similar to those at zero field in Fig.~\ref{fig:spin}(e) but has the definite sign of $n_{\rm sk}$ in the magnetic field. 

While further increasing $H^z$, the skyrmion number changes from $-1$ to $0$ at $H^z \simeq 1.7$, as denoted by the dashed line in Fig.~\ref{fig:Hz}(a), although the other spin and chirality quantities continuously change.  
We show the spin and chirality configurations below ($H^z=1.5$) and above ($H^z=2$) the transition in Figs.~\ref{fig:spin_Hz}(c) and \ref{fig:spin_Hz}(d), respectively. 
One can find that the sign of $S_i^z$ at the skyrmion core is reversed in the transition, and accordingly, the chirality around the core takes an opposite sign as well. 
The latter state is regarded as the triangular vortex crystal, which has been discussed in the context of Mott insulators~\cite{Kamiya_PhysRevX.4.011023,Wang_PhysRevLett.115.107201,Hayami_PhysRevB.94.174420} and metals~\cite{hayami2020multiple,Hayami_PhysRevB.103.054422,yambe2021skyrmion}. 
Further increment of $H^z$ gradually changes the spin texture so as to have more $+z$-spin component, as shown in Fig.~\ref{fig:spin_Hz}(e). 
Thus, the scalar chirality is suppressed against the field in this state.
Finally, the 3$Q$ state turns into the fully-polarized state at $H^z \simeq 4$. 

\subsection{Field along the $x$ direction}
\label{sec:x}

\begin{figure}[t!]
\begin{center}
\includegraphics[width=0.9\hsize]{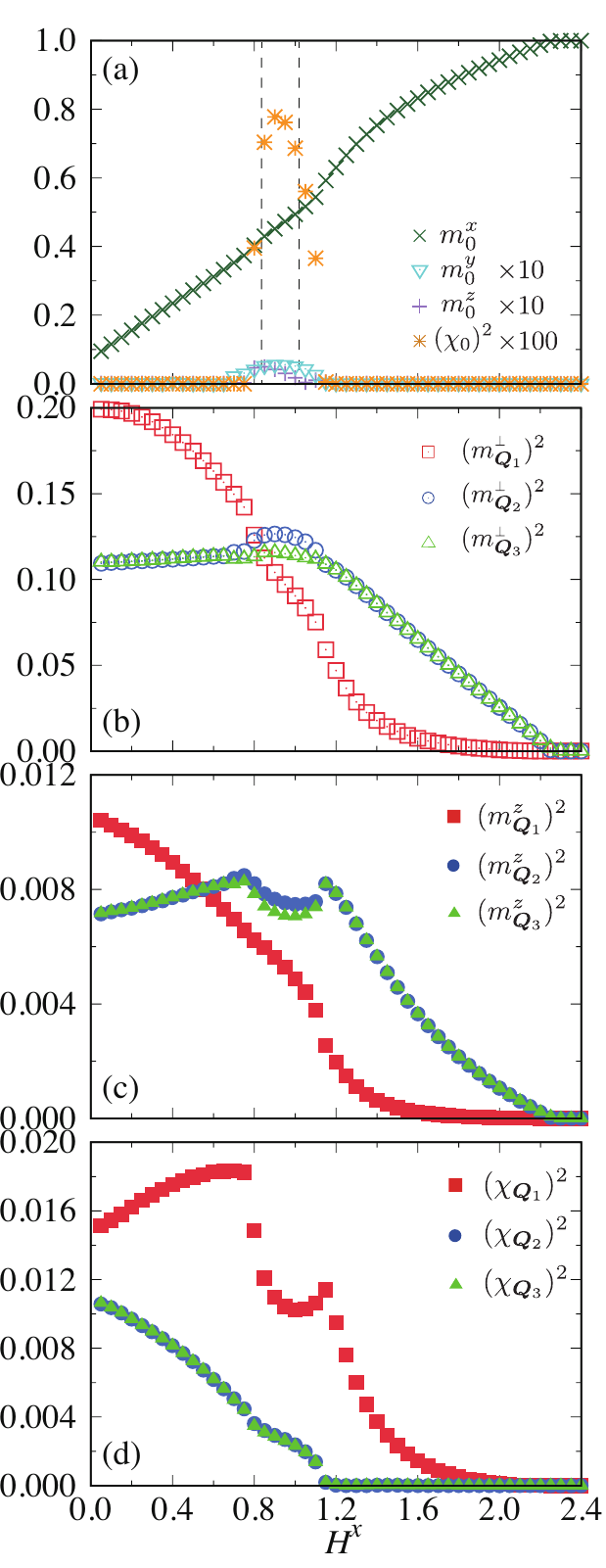} 
\caption{
\label{fig:Hx}
$H^x$ dependence of (a) $\bm{m}_0=(m^x_0,m^y_0,m^z_0)$ and $(\chi_0)^2$, (b) $(m^{\perp}_{\bm{Q}_\nu})^2$, (c) $(m^{z}_{\bm{Q}_\nu})^2$, and (d) $ (\chi_{\bm{Q}_\nu})^2$ for $D=0.2$ and $K=0.6$.
The dashed line in (a) represents the magnetic field where the behavior the skyrmion number changes (see the text in details).
}
\end{center}
\end{figure}

\begin{figure}[t!]
\begin{center}
\includegraphics[width=1.0\hsize]{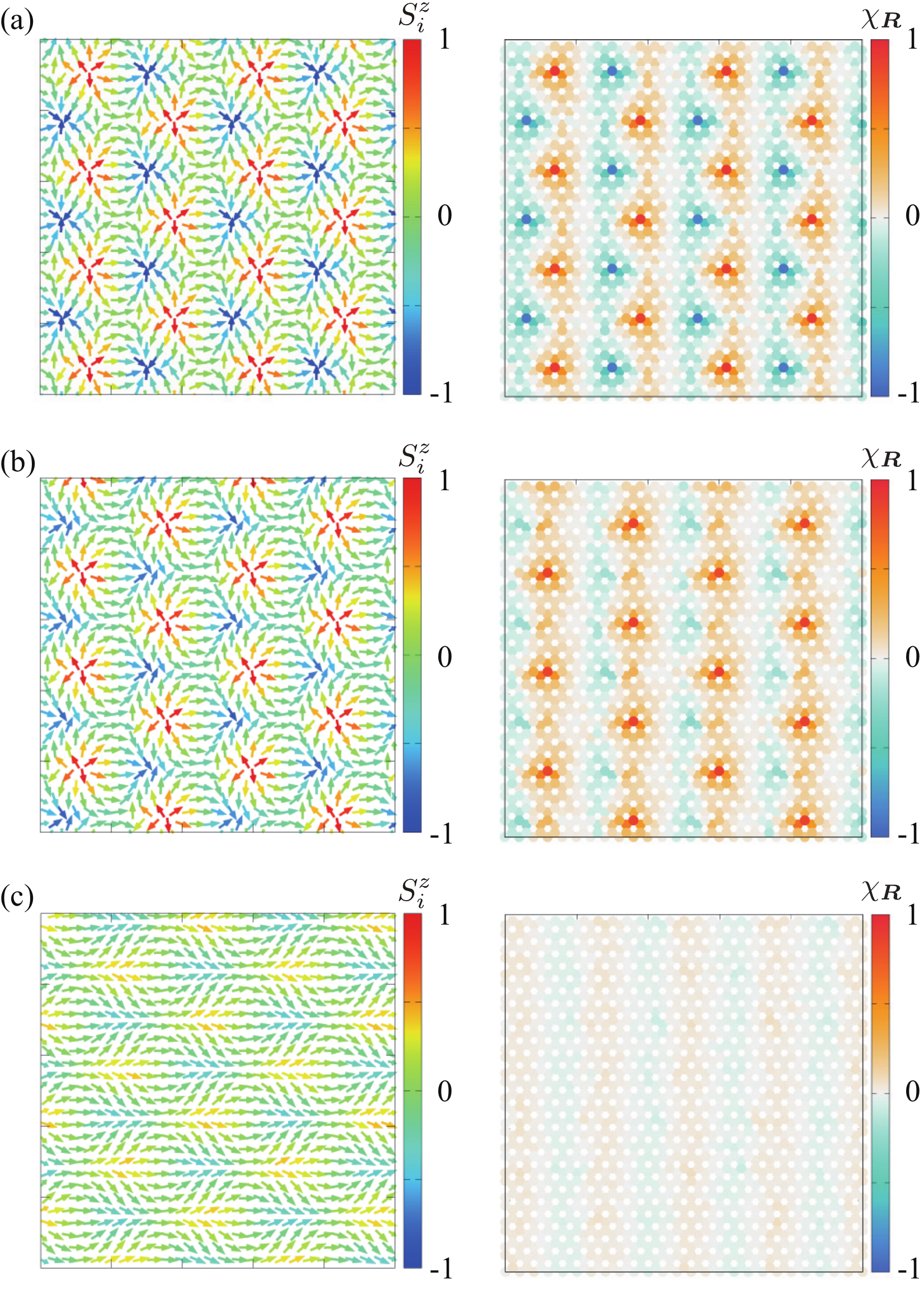} 
\caption{
\label{fig:spin_Hx}
 (Left panel) Snapshots of the spin configurations for (a) $H^x=0.5$, (b) $H^x=1$, and (c) $H^x=1.8$ at $D=0.2$ and $K=0.6$. 
The contours show the $z$ component of the spin moment, while the arrows represent the $xy$ components of the spin moment. 
(Right panel) Snapshots of the scalar chirality configurations. 
}
\end{center}
\end{figure}

We consider the effect of the magnetic field along the $x$ direction, $\bm{H}=(H^x,0,0)$. 
Figure~\ref{fig:Hx} shows $H^x$ dependences of spin and chirality with $\bm{q}=\bm{0}, \bm{Q}_1$, $\bm{Q}_2$ and $\bm{Q}_3$. 
For nonzero $H^x$, the spin moments are rearranged so as to make $\bm{m}_{\bm{Q}_1}$ large compared to $\bm{m}_{\bm{Q}_2}$ and $\bm{m}_{\bm{Q}_3}$. 
In the presence of the field, $\bm{m}_{\bm{Q}_2}$ is equivalent to $\bm{m}_{\bm{Q}_3}$, which results in the uniform magnetization along the $x$ direction and no scalar chirality.
The spin and chirality configurations at $H^x=0.5$ are shown in Fig.~\ref{fig:spin_Hx}(a). 
From the chirality configuration, one finds that the stripy pattern of $\chi_{\bm{R}}$ is dominant along the $\bm{Q}_1$ direction [see also Fig.~\ref{fig:Hx}(d)], which is different from the almost threefold rotational symmetric pattern in the case of $H^z$ in Fig.~\ref{fig:spin_Hz}(a). 

While increasing $H^x$, we find that the system exhibits nonzero $\chi_0$ in the intermediate field for $0.8 \lesssim H^x \lesssim 1.1$. 
In this state, the amplitude of $\bm{m}_{\bm{Q}_2}$ is different from that of $\bm{m}_{\bm{Q}_3}$, as shown in Figs.~\ref{fig:Hx}(b) and \ref{fig:Hx}(c). 
This imbalance between $\bm{m}_{\bm{Q}_2}$ and $\bm{m}_{\bm{Q}_3}$ results in nonzero $m^y_0$ and $m^z_0$ as well as $\chi_0$ in Fig.~\ref{fig:Hx}(a). 
It is noted that both states with positive and negative uniform scalar chirality are obtained in the simulations depending on the initial spin configurations, which indicates that they are degenerate. 
By computing the skyrmion number in this state, it takes a nonzero and non-integer value around $0\lesssim |n_{\rm sk}| \lesssim 0.5$ for $0.825 \lesssim H^x \lesssim 1.025$ sandwiched by the dashed lines in Fig.~\ref{fig:Hx}(a), while it becomes zero in the other field region. The non-quantized skyrmion number might be ascribed to the short-period spin texture and/or thermal fluctuations, which needs a more careful analysis. 
Figure~\ref{fig:spin_Hx}(b) shows the spin and chirality configurations at $H^x=1$. 
In the snapshot, the spin configuration has a positive net chirality and a positive (negative) magnetization along the $y$ ($z$) direction. 

For $H^x \gtrsim 1.1$, $\chi_0$, $m^{\perp}_{\bm{Q}_1}$, $m^{z}_{\bm{Q}_1}$, $\chi_{\bm{Q}_2}$, and $\chi_{\bm{Q}_3}$ are suppressed (Fig.~\ref{fig:Hx}), and as a result, the state becomes a topologically trivial 3$Q$ state with the chirality density wave along the $\bm{Q}_1$ direction. 
The behavior that $m^{\perp}_{\bm{Q}_1}$ ($m^{z}_{\bm{Q}_1}$)  is rapidly suppressed compared to $m^{\perp}_{\bm{Q}_2}$ ($m^{z}_{\bm{Q}_2}$) and $m^{\perp}_{\bm{Q}_3}$ ($m^{z}_{\bm{Q}_3}$) is reasonable, since the $x$ directional field favors the conical spiral perpendicular to the field direction in polar magnets. 
The spin and scalar chirality configurations in the 3$Q$ state is presented in Fig.~\ref{fig:spin_Hx}(c).
This 3$Q$ state continuously turns into the fully-polarized state at $H^x \simeq 2.25$.

\subsection{Field along the $y$ direction}
\label{sec:y}

\begin{figure}[t!]
\begin{center}
\includegraphics[width=0.92\hsize]{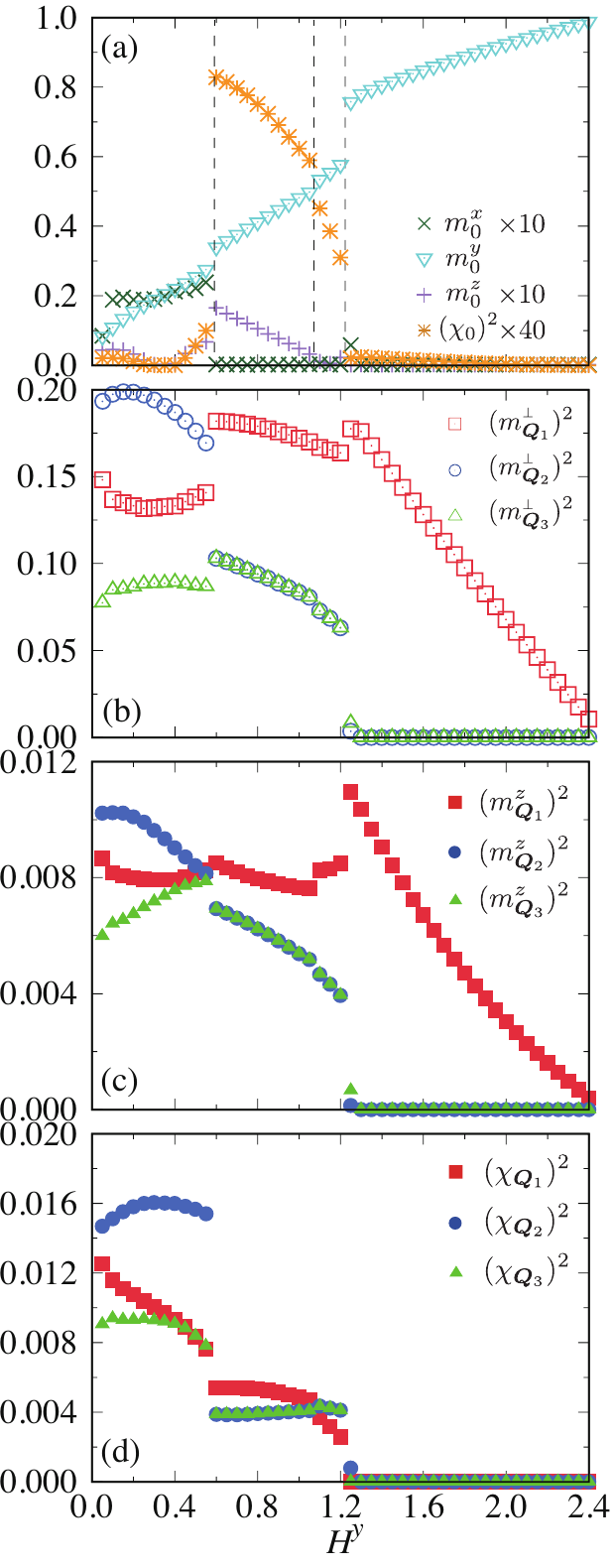} 
\caption{
\label{fig:Hy}
$H^y$ dependence of (a) $\bm{m}_0=(m^x_0,m^y_0,m^z_0)$ and $(\chi_0)^2$, (b) $(m^{\perp}_{\bm{Q}_\nu})^2$, (c) $(m^{z}_{\bm{Q}_\nu})^2$, and (d) $ (\chi_{\bm{Q}_\nu})^2$ for $D=0.2$ and $K=0.6$.
The dashed line in (a) represents the magnetic field where the behavior the skyrmion number changes (see the text in details). 
}
\end{center}
\end{figure}

\begin{figure}[t!]
\begin{center}
\includegraphics[width=1.0\hsize]{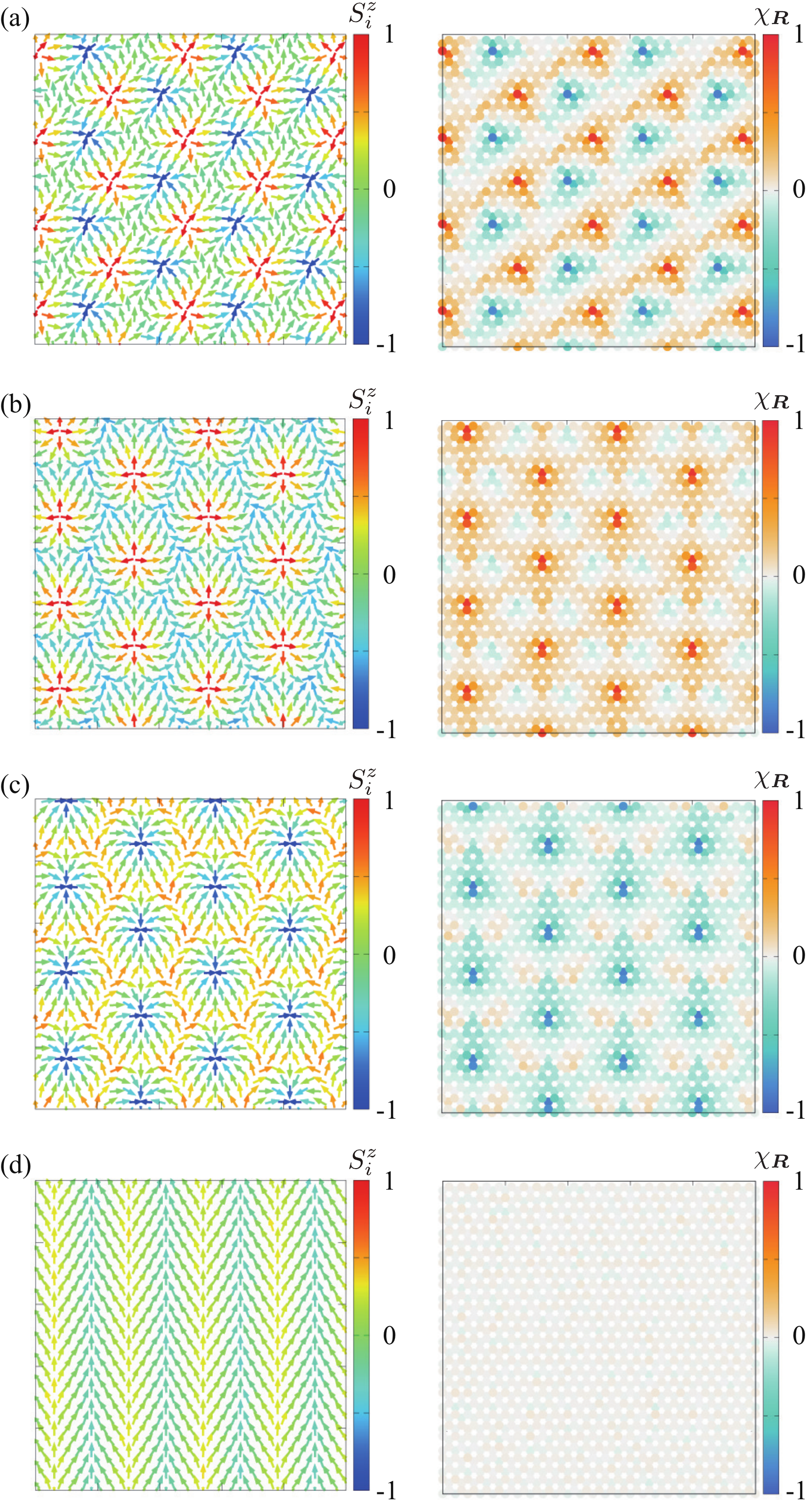} 
\caption{
\label{fig:spin_Hy}
 (Left panel) Snapshots of the spin configurations for (a) $H^y=0.5$, (b) $H^y=0.7$, (c) $H^y=0.8$, and (d) $H^y=1.8$ at $D=0.2$ and $K=0.6$. 
The contours show the $z$ component of the spin moment, while the arrows represent the $xy$ components of the spin moment. 
(Right panel) Snapshots of the scalar chirality configurations. 
}
\end{center}
\end{figure}

We discuss the result when applying the magnetic field along the $y$ direction, $\bm{H}=(0,H^y,0)$.  
Notably, the $y$-field effect on the 3$Q'$ MAX is different from those along the $x$ and $z$ directions. 
When $H^y$ is applied to the system, the $3Q'$ MAX exhibits a nonzero $\chi_0$, as shown in Fig.~\ref{fig:Hy}(a). 
Besides, the uniform magnetizations along the $x$ and $z$ directions perpendicular to the field direction are also induced. 
While increasing $H^y$, $\chi_0$ shows a nonmonotonic behavior; $\chi_0$ decreases for $H^y \lesssim 0.3$ and increases for $0.3 \lesssim H^y \lesssim 0.6$. 
This spin state is characterized by the anisotropic 3$Q$ spin texture with $\bm{m}_{\bm{Q}_2}>\bm{m}_{\bm{Q}_1}>\bm{m}_{\bm{Q}_3}$. 
The spin and chirality configurations at $H^y=0.5$ are shown in Fig.~\ref{fig:spin_Hy}(a). 
In the state for $0 < H^y \lesssim 0.6$, $|n_{\rm sk}|$ takes a nonzero and non-integer value around $0\lesssim |n_{\rm sk}| \lesssim 0.95$, similar to the case along the $x$ field direction in Sec.~\ref{sec:x}; $n_{\rm sk}$ takes close to one for the small field for $0<H^y \lesssim 0.15$, but close to zero for the other region. 

For $H^y \simeq 0.6$, the low-field state turns into the other chiral state with a jump of $\chi_0$ and the other quantities, as shown in Fig.~\ref{fig:Hy}. 
The value of $\chi_0$ is largely enhanced and $m_0^x$ vanishes shown in Fig.~\ref{fig:Hy}(a). 
Simultaneously, the amplitude of $\bm{m}_{\bm{Q}_2}$ becomes equivalent to $\bm{m}_{\bm{Q}_3}$. 
The snapshots of the spin and scalar chirality in this state are shown in Fig.~\ref{fig:spin_Hy}(b) in the case of $H^y = 0.7$. 
From the real-space spin and chirality configurations, one finds that vortices with $S_i^z \simeq +1$ form the almost triangular lattice, which resembles the situation in the SkX under the magnetic field along the $z$ direction. 
Indeed, we obtain the quantized skyrmion number of $+1$ in this state. 
Thus, the SkX also appears in the intermediate field along the $y$ direction in addition to the case along the $z$ direction. 
It is noted that the SkX with the opposite sign of the net chirality is also obtained depending on the initial spin configurations. 
We show the spin and chirality textures in the SkX with $n_{\rm sk}=-1$ obtained by simulated annealing at $H^y=0.8$ in Fig.~\ref{fig:spin_Hy}(c). 

While further increasing $H^y$, the SkX is replaced by the other chiral state with nonzero $\chi_0$ but non-quantized $n_{\rm sk}$ for $1.1 \lesssim H^y \lesssim 1.2$, where $0\lesssim |n_{\rm sk}| \lesssim 0.95$. 
Then, this state turns into the 1$Q$ conical state with nonzero $\bm{m}_{\bm{Q}_1}$ around $H^y\simeq 1.25$, as shown in Figs.~\ref{fig:Hy}(b) and \ref{fig:Hy}(c). 
The finite $\chi_0$ for $H^y \gtrsim 1.25$ in Fig.~\ref{fig:Hy}(a) is owing to the staggered contribution of the chirality for upward and downward triangles, whose summation becomes zero, i.e., no net scalar chirality.
$\chi_{\bm{Q}_\nu}$ becomes zero in the 1$Q$ conical state, as shown in Fig.~\ref{fig:Hy}(d). 
The spin configuration of the 1$Q$ conical state is presented in Fig.~\ref{fig:spin_Hy}(d). 
Finally, the 1$Q$ conical state continuously turns into the fully-polarized state for $H^y \simeq 2.45$.

\section{Summary}
\label{sec:Summary}

To summarize, we investigate the stability of the MAX in noncentrosymmetric itinerant magnets on the triangular lattice by focusing on the interplay between the spin-charge coupling and the spin-orbit coupling.  
By performing simulated annealing for the effective spin model with the long-range interactions in momentum space, we show that the model exhibits the rectangular (triangular)-shaped MAXs consisting of the 2$Q$ (3$Q$) cycloidal waves depending on the model parameters. 
In particular, we obtain three key ingredients for the stabilization of the triangular MAX: the biquadratic interaction, the DM interaction, and easy-plane single-ion anisotropy. 
We also examine the effect of the magnetic field on the triangular MAX by changing the field direction. 
We find that the triangular MAX shows highly anisotropic phase transitions; the MAX turns into the SkX in the cases of the $y$- and $z$-directional-field cases, while it does not for the $x$-field case. 
Our result indicates that noncentrosymmetric itinerant magnets with easy-plane magnetic anisotropy are a promising system for the stabilization of the MAXs, which would be a useful reference for further exploration of MAX-hosting materials

\begin{acknowledgments}
This research was supported by JSPS KAKENHI Grants Numbers JP19K03752, JP19H01834, JP21H01037, and by JST PRESTO (JPMJPR20L8). 
Parts of the numerical calculations were performed in the supercomputing systems in ISSP, the University of Tokyo.
\end{acknowledgments}

\bibliographystyle{apsrev}
\bibliography{ref}

\end{document}